\documentclass[11pt,graphicx,amsmath]{article}
\usepackage{amsmath}
\usepackage{graphicx}
\usepackage{bm}
\usepackage{color}
\usepackage{amssymb}
\usepackage{amsfonts}
\usepackage{comment}
\usepackage{cite}
\usepackage{todonotes}
\usepackage{caption}
\usepackage{subcaption}
\usepackage{appendix}
\usepackage{epstopdf}

\def\be{\begin{equation}}
\def\ee{\end{equation}}

\def\ba{\begin{eqnarray}}
\def\ea{\end{eqnarray}}
\def\bl#1\el{\begin{align}#1\end{align}}

\def\bl#1\el{\begin{align}#1\end{align}}

\title{Field Equation of Correlation Function of
Mass Density Fluctuation for Self-Gravitating Systems }

 \author{\small
            \,  Yang  Zhang\thanks{yzh@ustc.edu.cn} , \,
             Qing Chen  \thanks{cqpb@mail.ustc.edu.cn}
             \\
 \small  Department of  Astronomy,
   CAS Key Laboratory for Researches in Galaxies and Cosmology, \\
 \small  University of Science and Technology of China, Hefei, Anhui, 230026, China \\
 }

 \date{}

\def\be{\begin{equation}}
\def\ee{\end{equation}}
\def\ba{\begin{eqnarray}}
\def\ea{\end{eqnarray}}

\newcommand{\bee}{\begin{eqnarray}}
\newcommand{\een}{\end{eqnarray}}

\begin{document}

\maketitle

\begin{abstract}\Large

We study the mass density distribution of
Newtonian self-gravitating systems.
Modeling the system  as a fluid in hydrostatical equilibrium,
we obtain from first principle
 the field equation and its solution
of  correlation function $\xi(r)$ of
the mass density fluctuation itself.
We apply thid  to studies of the large-scale structure of
the Universe within a small redshift range.

The equation tells that
$\xi(r)$ depends on the point mass $m$ and  the Jeans wavelength scale $\lambda_{0}$,
which are different for galaxies and clusters.
It explains several longstanding, prominent features
of the observed clustering :
that the profile of $\xi_{cc}(r)$ of clusters is similar to $\xi_{gg}(r)$
of galaxies but with a higher amplitude and a longer correlation length,
and that the correlation length increases with the mean separation between
clusters as a universal scaling $r_0\simeq 0.4d$.
Our  solution $\xi(r)$ also yields
the observed power-law  correlation function of galaxies
$\xi_{gg}(r)\simeq  (r_0/r)^{1.7}$
valid only in a range $1<r<10 h^{-1}$Mpc.
At larger scales the solution $\xi(r)$ breaks below the power law
and goes to zero around $\sim 50h^{-1}$Mpc,
just as the observational data have demonstrated.
With a set of fixed  model parameters,
the solutions $\xi_{gg}(r)$ for galaxies,
 the corresponding power spectrum,  and $\xi_{cc}(r)$ for clusters,
simultaneously\textbf{,}
 agree with the observational data
from the major surveys of galaxies, and of clusters.

\end{abstract}


Published in AA 581, A53 (2015),  doi: 10.1051/0004 -6361/201425431

\large

\section{Introduction}

It is
one of the major goals of modern cosmology
to understand the  matter distribution in the Universe on large scales.
The large-scale structure is determined by self gravity of galaxies and clusters.
Since the number of galaxies  is enormous,
one needs statistics to study their distribution.
In this regard,
the $2$-point correlation functions $\xi_{gg}(r)$ of galaxies
and $\xi_{cc}(r)$ of clusters
serve as a powerful statistical tool \cite{Bok,TotsujiKihara,Peebles}.
It  not only provides  statistical information,
but also contains underlying dynamics mainly due to gravitational force.
Therefore, we would like to investigate the correlation functions of
self gravitating systems in approximation of hydrostatical equilibrium.

Over the years,
various observational surveys have been carried out
for galaxies and for clusters,
such as the Automatic Plate Measuring (APM) galaxy survey
                   \cite{loveday1996stromlo},
the Two-degree-Field Galaxy Redshift Survey (2dFGRS)\cite{Peacock2001},
Sloan Digital Sky Survey (SDSS)\cite{abazajian2009seventh}, etc.
All these surveys suggest that the correlation of galaxies
has a power-law form
$\xi_{gg}(r)\propto (r_0/r)^{\gamma}$ with $r_0\sim 5.4h^{-1}$Mpc
and $\gamma\sim 1.7$ in a range $ (0.1\sim 10)h^{-1}$Mpc
\cite{TotsujiKihara,groth1977large,Peebles,SoneraPeebles}.
The correlation of clusters is found to be
of a similar form:  $\xi_{cc}(r)\sim 20 \xi_{gg}(r)$
 in a range $ (5\sim 60)h^{-1}$Mpc,
with an amplified magnitude \cite{1983ApJ...270...20B,KlypinKopylov}.
For quasars, the correlation is $\xi_{qq}(r)\sim 5 \xi_{gg}(r)$ \cite{Shaver}.
Numerical computations
have been extensively employed to study  clustering of galaxies and of clusters,
and significant progresses have been made.
To understand  physical mechanisms  behind the clustering,
 analytical studies are important.
In particular,
references \cite{Saslaw85,Saslaw2000} used  thermodynamics
whereby the power-law form of $\xi_{gg}(r)$
was introduced as modifications to the energy and pressure.
Similarly, references \cite{deVegaSanchez,deVegaSanchez98}  used
the grand partition function of \textbf{a} self-gravitating gas
to study a possible fractal structure of
the distribution of galaxies.
However \textbf{,} the field equation of $\xi$
was not given in these studies.
In this paper we use units that the speed of light is $c=1$
 and the Boltzmann constant is $k_B=1$.

\section{Field equation of the 2-pt correlation function
of density fluctuations}

Galaxies, or clusters, distributed in the Universe
can be described as fluids at rest in   gravitational fields.
This modeling is  an approximation
since the cosmic expansion is not considered.
We apply   hydrostatics
to   systems of galaxies within a small redshift range.
For these  self-gravitating systems,
the  field equation of mass density is \cite{Zhang2007,zhang2009nonlinear}
\begin{equation}  \label{field_eq}
\nabla^{2}\psi-\frac{1}{\psi}(\nabla\psi)^{2}+k_{J}^{2}\psi^{2}+J\psi^{2}=0\textbf{,}
\end{equation}
where $\psi ({\bf r}) \equiv \rho({\bf r})/\rho_{0}$
with $\rho_0=mn_0$ being the mean mass density of the system,
$k_{J}\equiv \sqrt{4\pi G\rho_{0}}/c_{s}$ is the Jeans wavenumber,
and $J$ is a  Schwinger  type  of  external source  introduced for
taking functional derivative  \cite{Schwinger}.
The effective Hamiltonian density \textbf{is}
\begin{equation}  \label{eff_L}
\mathcal{H}(\psi,J)=\frac{1}{2}(\frac{\nabla\psi}{\psi})^{2}
               -k_{J}^{2}\psi-J\psi\textbf{.}
\end{equation}
The generating functional for
the correlation functions of $\psi$ is
\begin{equation} \label{ZJ}
Z[J] = \int  D\psi
  e^{-\alpha \int d^{3}\textbf{r}\mathcal{H}(\psi,J)},
\end{equation}
where $\alpha  \equiv c_s^2/4\pi G m$,  $c_s $ is the sound speed
and $m$ is the mass of a single particle.

Since the distribution of galaxies, or clusters,
can be viewed as fluctuations of the mass density in the Universe,
we consider the fluctuation field
$\delta\psi(\bf{r}) \equiv  \psi(\textbf{r})-\langle\psi(\textbf{r})\rangle$,
where the statistical ensemble  average is defined as
$ \langle\psi(\textbf{r})\rangle =
  \frac{\delta}{\alpha \delta J({\bf r })}\log Z[J] \mid_{J=0} $,
and, in our case, $\langle\psi(\textbf{r})\rangle=\psi_{0}$ is a constant.
The {\it connected} n-point correlation function of $\delta \psi$ is defined
as \cite{Binney}
\begin{eqnarray}
G^{(n)}({\bf r}_{1},..., {\bf r}_{n})
&\equiv \langle\delta\psi({\bf r}_{1})...\delta\psi({\bf r}_{n}) \rangle\nonumber\\
& =\alpha^{-(n-1)}
 \frac{\delta^{n-1} \langle\psi({\bf r}_n)\rangle_J}{\delta J({\bf r}_1)
       ...\delta J({\bf r}_{n-1} )}   |_{J=0}
\end{eqnarray}
for $n\ge 2$.
One can take
$G^{(2)}({\bf r}_{1},{\bf r}_{2}) = G^{(2)}(r_{12}) $
for the  homogeneous and isotropic Universe.
To derive the field equation of $G^{(2)}(r)$  \cite{Goldenfeld},
one takes functional derivative of the ensemble average of Eq. \ref{field_eq}
with respect to $J({\bf r}_1)$.
In doing this,   $G^{(3)}$ occurs in the equation of  $G^{(2)}(r)$
hierarchically.
We adopt the Kirkwood-Groth-Peebles  ansatz \cite{Kirkwood,groth1977large}
$
G^{(3)}(\textbf{r}_1,\textbf{r}_2,\textbf{r}_3)
=Q[G^{(2)}(r_{12} ) G^{(2)}(r_{23} )
 +G^{(2)}(r_{23} ) G^{(2)}(r_{31} )
 +G^{(2)}(r_{31} ) G^{(2)}(r_{12} )],
$
where $Q$ is  a dimensionless parameter.
Then,  after a necessary renormalization,
we obtain the field equation of the 2-point correlation function
\begin{equation} \label{final}
(1-b\xi)\xi''+
( (1-b\xi)\frac{2}{x}+a )\xi' +\xi  -b \xi'\, ^{2}   -c\xi^{2}=
-\frac{1}{\alpha }   \frac{\delta(x)k_0}{x^{2}},
\end{equation}
where  $\xi=\xi(r) \equiv  G^{(2)}(\textbf{r})$,
$\xi'\equiv \frac{d}{dx}\xi$,  $x \equiv k_{0}r$,
$k_0\equiv \sqrt{2}k_J$,
and  $ a$, $b$, and $c$ are three independent parameters.
Eq. \ref{final}  extends that in the earlier work \cite{zhang2009nonlinear}.
The special case of $ a=b=c=0$  is the Gaussian approximation.
The terms containing $ a$, $b$, and $c$
represent the nonlinear contributions beyond
the Gaussian approximation.
The nonlinear terms with $b$ and $c$ in Eq. \ref{final}
can enhance the amplitude of $\xi$ at small scales
and increase the correlation length.
The term containing  $a$ plays \textbf{a} role of effective viscosity.
 The value of $a$ should be large enough
to ensure  $1+\xi(r)  \ge 0 $ for the whole range  $0<r<\infty$.

\section{General predictions of  field equation}

We inspect Eq. \ref{final} to
see its predictions on general properties of the correlation function $\xi(r)$.

1, The  equation contains a point mass $m$ and a characteristic wavenumber $k_0 $.
It applies to the system of galaxies,
as well as to the system of clusters,
but with different $m$ and $k_0$ in each respective case.
Thus, as  solutions of Eq. \ref{final},
$\xi_{cc}$  for  clusters should have a profile similar to $\xi_{gg}$ for galaxies,
but will differ in amplitude and in scale
determined by different $m$ and $k_0$.
Indeed, the observations show  that both
 $\xi_{gg}$   and $\xi_{cc}$
have a power-law form: $\propto r^{-1.8}$ in their respective, finite range,
but  $\xi_{cc}$ has a higher amplitude \cite{1983ApJ...270...20B,KlypinKopylov}.

2,  The $\delta^{3}(\bf r)$ source in Eq. \ref{final}
has a coefficient $ 1/\alpha =4\pi G m /c_s^2$,
which determines the overall amplitude of a solution $\xi$.
The mass $m$ of a cluster can be $10 \sim 10^3$ times that of a galaxy \cite{Bahcall1999},
while $c_s$ regarded as the the peculiar velocity
is  around several hundreds km/s for galaxies and clusters.
Therefore, $1/\alpha \propto m$,
and a greater $m$ will yield a higher amplitude of $\xi$.
This general prediction
naturally explains a whole chain of prominent facts of observations,
that luminous galaxies are more massive and have a higher correlation amplitude
  than ordinary galaxies \cite{Zehavi05},
  that clusters are much more massive and have a much higher correlation than galaxies,
 and that rich clusters have a higher correlation
         than poor clusters
         since richness is proportional to  mass
\cite{1983ApJ...270...20B,Einasto02,Einasto2000b,Bachall03}.
        These phenomena have been a puzzle for long \cite{Bahcall1999}
         and were interpreted as being caused by the statistics
         of rare peak events \cite{Kaiser1984}.

3, The power spectrum
is proportional to the inverse of the spatial number density:
$P(k) \propto 1/n_0$.
Given the mean mass density $\rho_0=m n_0$,
a greater $m$ implies a lower $n_0$.
Therefore, the properties  2 and 3
reflect the same physical law of clustering from different perspectives.
The property 3 also agrees with the observed  fact
from a variety of surveys.
The observed $P(k)$ of clusters is much higher than that of galaxies,
and the observed $P(k)$ of rich clusters is higher than poor clusters, etc.
This is because that $n_0$ of clusters is much lower than that of galaxies,
and $n_0$ of rich clusters is lower than that of poor clusters
\cite{Bahcall1999}.

4,   The characteristic length
$\lambda_0=2\pi/k_0  \propto \frac{c_s}{\sqrt{\rho_0}}$
appears in Eq. \ref{final} as the only scale
that underlies the scale-related features of clustering.
Cluster surveys extend over larger spatial volumes,
including those very dilute regions.
The mass density  $\rho_{0c}$ of the region covered by cluster surveys
is lower than  $\rho_{0g}$ for galaxy surveys,
  as implied in the references  \cite{Bahcall1999,BahcallLubin}.
Thus,  $\lambda_0 $ for cluster surveys will be longer than that for galaxy surveys.
As will be seen in  Sects. 4 and 5,
in using one solution $\xi(r)$
to match the data of both galaxies and clusters,
one needs  to take $k_0$  smaller for clusters
than for galaxies.

\section{ Confronting   observational data of galaxy surveys}

Now we give the solution $\xi_{gg}(r)$ for a fixed set of parameters
$(a,b,c)$,
and \textbf{confront the} observed correlation from major galaxy surveys.

1, The correlation function $\xi_{gg}(r)$

\begin{figure}
\includegraphics[width=\linewidth]{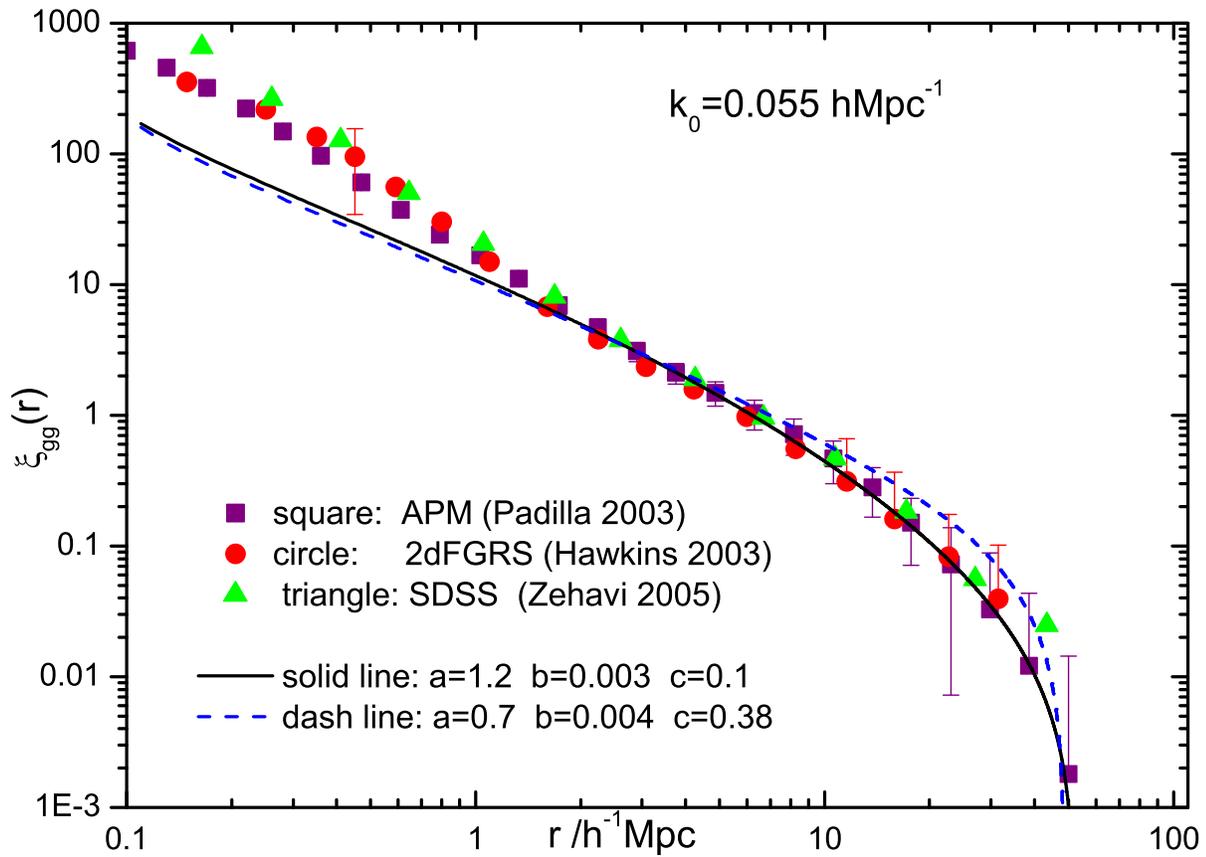}
\caption{
The solution $\xi_{gg}(r)$ confronts the data of galaxies by
 APM \cite{Padilla03},
 2dFGRS \cite{hawkins20032df}, and SDSS \cite{Zehavi05}.
 Here $k_0=0.055$ hMpc$^{-1}$ is taken in calculation.
}
\label{correlation}
\end{figure}

Fig. \ref{correlation} shows
the solution $\xi_{gg}(r)$ and the observational  data by the galaxy surveys
of APM \cite{Padilla03}, SDSS \cite{Zehavi05}, and 2dFGRS \cite{hawkins20032df}.
The theoretical $\xi_{gg}(r)$
matches the  data
 in the range of $r=(2 \sim 50)$ h$^{-1}$Mpc.
The usual power-law  $\xi_{gg} \propto r^{-1.7}$ is valid only
 in an interval  $(0.1 \sim 10)$ h$^{-1}$Mpc.
On large scales, the solution $\xi_{gg}(r)$ deviates from the power law,
decreases rapidly to zero,  and becomes negative around $50$ h$^{-1}$Mpc.
However, on small scales $r \leq 1$ h$^{-1}$Mpc,
 the solution  $\xi_{gg}(r)$ is lower than the data,
even though it has already improved the Gaussian approximation\cite{Zhang2007}.
This insufficiency at $r \leq 1$ h$^{-1}$Mpc
should be due to neglect of high order
nonlinear terms such as  $(\delta \psi)^3$
 in our perturbation.
We remark that the equation of
$\xi_{gg}(r) $ has been derived assuming  $\delta\psi <1$.
Thus, it is only an approximation to extrapolate
our calculated  $\xi_{gg}(r)$ down  to  smaller  scales  $r \leq 5 h^{-1}$Mpc.

Let us check that  the approximation  of hydrostatical equilibrium  can be applied
in the quasi-linear regime in the expanding Universe.
The density fluctuation behaves as $\delta\psi  \propto a(t)^{0.3} $ approximately,
where   $a(t)$ is the scale factor in the present stage of accelerating expansion.
So the   time-evolving correlation function
$\xi_{gg}(r,t)=
\langle \delta\psi  \delta\psi  \rangle \propto a^{0.6}(t)=1/(1+z)^{0.6}$.
For the  sample of $\sim 200,000$ galaxies of SDSS \cite{Zehavi05},
the redshift range is $z=(0.02\sim 0.167)$.
Taking its maximum $z = 0.167 $ into
the ratio gives  $\xi_{gg}(r)/\xi_{gg}(r,t) \simeq  (1+0.167)^{0.6} \sim 1.097$,
and the  error  \textbf{is} $0.6z \simeq  0.1 $.
The conclusion of this analysis has  also been supported
  by studies of  numerical simulations
\cite{Hamana,Yoshikawa,Taruya2001}.

2, The power spectrum $P(k)$

\begin{figure}
\includegraphics[width=\linewidth]{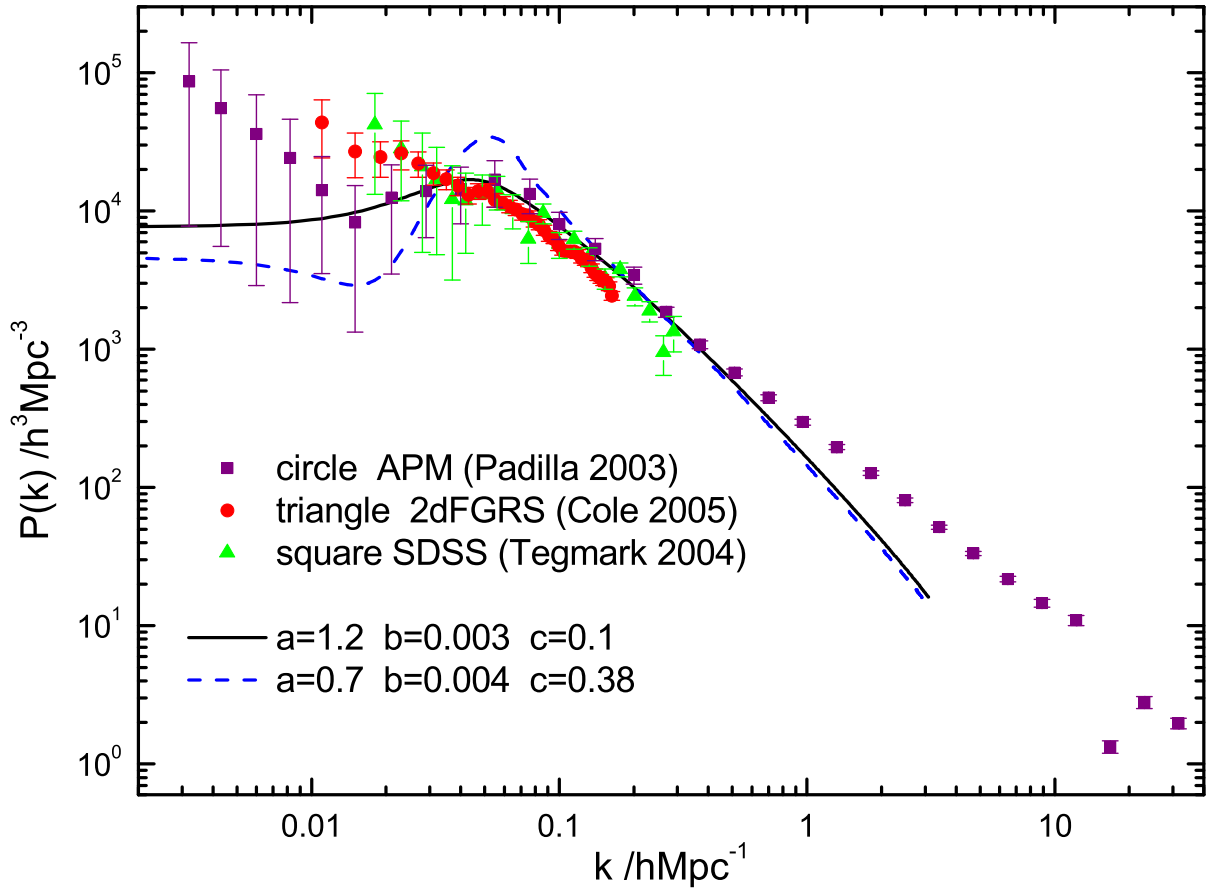}
\caption{
The power spectrum $P(k)$ converted
from  $\xi_{gg}(r)$ in Fig.  \ref{correlation}
confronts the data of  APM \cite{Padilla03}, 2dFGRS \cite{cole20052df}
and SDSS \cite{blantonTegmark2004}.
}
\label{Pk}
\end{figure}

The power spectrum $P(k)$ is the Fourier transform
\begin{equation} \label{P(k)}
P(k)=4\pi\int\limits_{0}^{\infty}\xi(r)\frac{\sin(kr)}{kr}r^{2}dr
\end{equation}
of the  correlation function $\xi(r)$.
It measures  the mass  density fluctuation in  $k$-space.
In principle, $P(k)$ and $\xi(r)$ contain the same information
if both are complete on their respective space,
$k=(0, \infty)$, and $r=(0, \infty)$.
Actually,  the observed $\xi_{gg}(r)$ is not complete,
limited to a finite range, say $r \leq 50$ Mpc.
If the observed power-law $\xi_{gg}(r)=(r_0/r)^{1.8}$ were plugged in Eq. \ref{P(k)},
one would have $P(k)\propto k^{-1.2}$,
which does not comply with the observed $P(k)\propto k^{-1.6}$ \cite{peacockbook}.
Our  $P(k)$ is obtained from the solution $\xi_{gg}(r)$
given on the whole range $r=(0,\infty)$.
Fig.  \ref{Pk} shows the theoretical $P(k)$ converted by Eq. \ref{P(k)}
from the solution  $\xi_{gg}(r)$
with the same set $(a,b,c)$ and $k_0$ as those in Fig. \ref{correlation}.
Also shown in Fig. \ref{Pk} are the observational data of $P(k)$
from APM \cite{Padilla03}, 2dFGRS \cite{cole20052df},
and SDSS \cite{blantonTegmark2004}.
It is seen that the theoretical $P(k)$ agrees well with the data $P(k)\propto k^{-1.6}$
 in the range of $k=(0.04\sim 0.7)$ hMpc$^{-1}$.
However, at large $k$,
the theoretical $P(k)$ is lower than the data.
This insufficiency of $P(k)$ corresponds to that of $\xi_{gg}(r)$
at small scales $r \leq 1h^{-1}$Mpc shown in Fig. \ref{correlation} .

\section{ Confronting the  observational data of clusters}

Clusters are believed to trace the cosmic mass distribution
on even larger scales,
and the observational data cover spatial scales farther than that of galaxies.
Now we shall apply
the solution with the same two sets of  $(a,b,c)$ as in \textbf{Sect. 4}
to the system of clusters.
A cluster has a mass $m$  greater than that of a galaxy.
This leads to a higher overall amplitude of $\xi_{cc}(r)$.
Besides, to match the observational data of clusters,
 a small value $k_0=0.03$ Mpc$^{-1}$ is required,
smaller than that for galaxies.
\begin{figure}
\includegraphics[width=\linewidth]{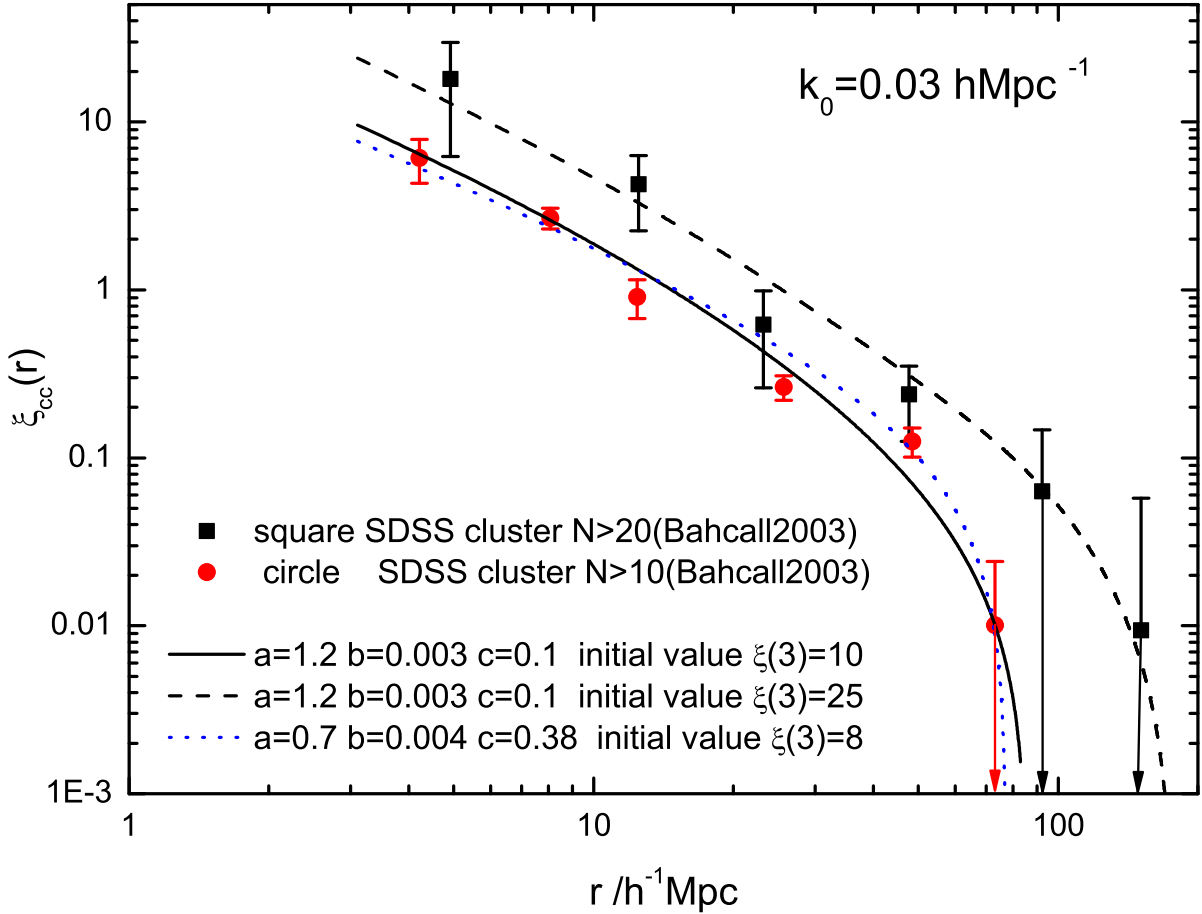}
\caption{The solution $\xi_{cc}(r)$ for clusters
confronts the data of SDSS clusters of two types of richness \cite{Bachall03}.
Here  $(a,b,c)$ are the same for galaxies.
But $k_0=0.03$ hMpc$^{-1}$ is taken for clusters,
 smaller than that for galaxies.}
\label{BahcallFig}
\end{figure}
In Fig. \ref{BahcallFig},
for each set of values of $(a,b,c)$ ,
two solutions $\xi_{cc}(r)$ with different amplitudes are given,
and are compared with two sets of data of richness $N> 10$ and  $N> 20$
from the SDSS  \cite{Bachall03}.
Interpreted by   Eq. \ref{final},
the $N> 20$  clusters have a greater $m$
than the $N> 10$ clusters.
The solutions match the data
on the whole range $r=(4\sim 100)$h$^{-1}$Mpc.

\begin{figure}
\includegraphics[width=\linewidth]{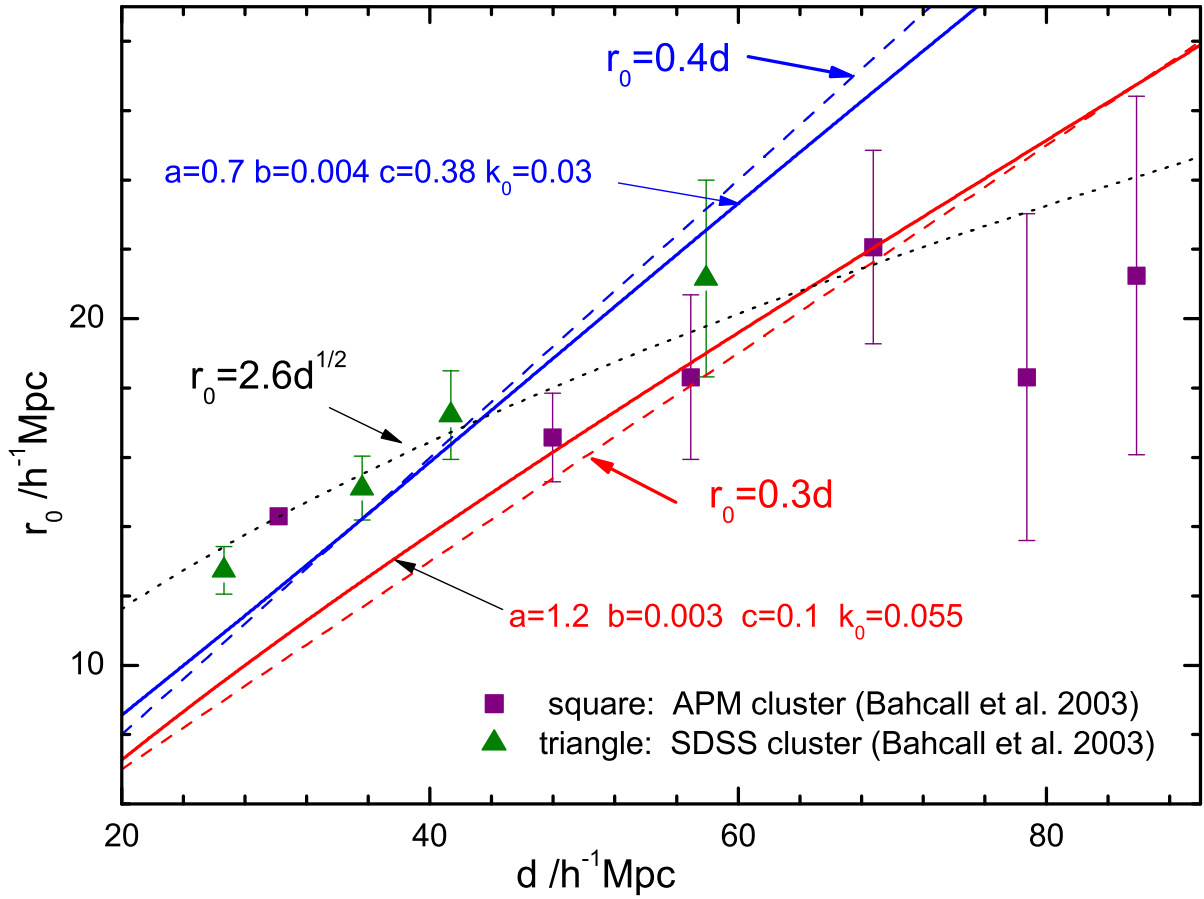}
\caption{The solution $\xi_{cc}(r)$ with $k_0=0.03$ hMpc$^{-1}$
gives the universal scaling  $r_0\simeq 0.4d$.
If a greater $k_0=0.055$ hMpc$^{-1}$ is taken,
$\xi_{cc}$ would give a flatter scaling $r_0\simeq 0.3d$,
which \textbf{fits} the data of APM clusters better \cite{Bachall03}.}
\label{scaling}
\end{figure}

It has long been known that there is a scaling behavior of  cluster correlation.
The correlation scale increases
with the mean spatial separation  between clusters
\cite{SzalaySchramm,BahcallWest1992,Bahcall1999,Croft, Gonzalez}.
If a power-law $\xi_{cc} =(r_0/r)^{1.8}$ was used to fit the data,
the  ``correlation length" would  be of a form
$r_0 \simeq 0.4d_i,$
where $d_i=n_i^{-1/3}$ and $n_i$
is the mean number density of clusters of type $i$.
Simulations have also produced this $r_0-d_i$ dependence \cite{BahcallCen}.
For SDSS, the scaling can also be fitted by
$r_0 \simeq 2.6 d_i\,^{1/2} $  \cite{Bachall03},
and for the 2df galaxy groups by $r_0\simeq 4.7 d_i\,^{0.32}$\cite{Zandivarez}.
This kind of universal scaling of  $r_0-d_i$
has been a theoretical challenge \textbf{\cite{Bahcall1997}},
and was thought to be either caused
by a fractal distribution of galaxies and clusters \cite{SzalaySchramm},
or by the statistics of rare peak events \cite{Kaiser1984}.
The  difference in the scaling slope was attributed to
different richness of clusters \cite{Bahcall1997}.
In our theory  the scaling is fully embodied in
the solution $\xi_{cc}(k_0 r)$
with the characteristic wavenumber $k_0=(8\pi G m n/c_s^2)^{1/2} \propto d^{-3/2}$.
To comply with the empirical power-law,
we take the theoretical ``correlation length" as
$r_0(d) \propto \xi_{cc}^{1/1.8}$,
where $\xi_{cc}$ is the theoretical solution and depends on  $d$.
Fig.  \ref{scaling} shows that
the solution $\xi_{cc}$ with $k_0=0.03$ hMpc$^{-1}$
gives the universal scaling  $r_0(d)\simeq 0.4d$,
agreeing well with the observation \cite{Bahcall1999}.
If  a greater $k_0=0.055$ hMpc$^{-1}$ is taken,
the solution $\xi_{cc}$ would yield a flatter scaling  $r_0(d)\simeq 0.3d$,
which fits the data of APM clusters better  \cite{Bachall03}.
Our analysis based on the solution $\xi_{cc}$ reveals that
a higher background density $\rho_0$
predicts  a flatter slope of the scaling  $r_0(d)$.
Conclusively,  the universal $r_0-d_i$ scaling is naturally
explained by the solution $\xi_{cc}(r)$.

\section{ Conclusions and discussions }

We have presented a field theory of density fluctuations of
Newtonian gravitating systems,
and applied it to the study of
correlation functions of galaxies and of clusters.
Starting from the field equation Eq. \ref{field_eq} of mass density,
under the condition of  hydrostatic equilibrium,
and keeping to the nonlinear order of $(\delta\psi)^2$ in perturbation,
by taking functional derivative,
we have obtained the field equation  Eq. \ref{final}
of the 2-point correlation function
as a main result.
In deriving Eq. \ref{final},
we have adopted the Kirkwood-Groth-Peebles ansatz
necessary to cut off the hierarchy in n-point correlation functions,
and have done renormalization
to absorb divergences into the parameters.
This analytic approach from first principle
is different from those using gravitational potential.
Eq.  \ref{final}  explains the observational data of clustering of
galaxies, and of clusters well on large scales.
To extend the analytic study further,
higher-order nonlinear terms should be included
to describe small-scale clusterings better,
and the cosmic evolution should be taken into consideration as
a more realistic model.

\section*{Acknowledgements}

Y. Zhang is supported by NSFC  No. 11275187, NSFC 11421303,
SRFDP, and CAS,
the Strategic Priority Research Program
"The Emergence of Cosmological Structures"
of the Chinese Academy of Sciences, Grant No. XDB09000000.


\begin{thebibliography}{99}
	

\bibitem{abazajian2009seventh}
{Abazajian, K. N., Adelman-McCarthy, J. K., Ag{\"u}eros, M. A.,  et al.
       2009, ApJS, 182, 543}

\bibitem{BahcallCen}     Bahcall, N.A., \& Cen, R.,
         1992, ApJ, 398, L81

\bibitem{BahcallWest1992}
                      {Bahcall, N.A., \& West, M.,
                      1992, ApJL, 392, 419}

\bibitem{1983ApJ...270...20B}
{Bahcall, N. A.,\& Soneira, R. M. 1983, ApJ, 270, 20}

\bibitem{BahcallLubin} {Bahcall, N.A.,  Lubin, L.M., \& Dorman, V.,
                   1995,  ApJ, 447,  L81}

\bibitem{Bahcall1997}  Bahcall, N.A.,
          1997, Large-Scale Structure in the Universe,
    in Unsolved Problems in Astrophysics, ed. by J.N. Bahcall and J.P. Ostriker,
            Princeton University Press.  arXiv:astro-ph/9612046.

\bibitem{Bahcall1999}  {Bahcall, N.A.,
          1999, Clusters and Supercluster,
in Formation of Structure in the Universe, ed. by A. Dekel and J.P. Ostriker,
Cambridge University Press.
 arXiv: astro-ph/9611148.}


\bibitem{Bachall03}
{Bahcall, N. A., Dong, F., Hao, L., et al. 2003, ApJ, 599, 814}



\bibitem{Binney}  {Binney, J. J, Dowrick, N., Fisher, A., \& Newman, M.,
        1992,  The Theory of Critical Phenomena,  Oxford University Press}



\bibitem{Bok} {Bok, B.J., 1934,  Bull Harvars Obs, 895,  1}

\bibitem{blantonTegmark2004}
{Blanton, M., Tegmark, M., \& Strauss, M., 2004, ApJ, 606, 702}



\bibitem{cole20052df}
{Cole, S., Percival, W. J., Peacock, J. A.,  et al., 2005, MNRAS, 362, 505}





\bibitem{Croft} {Croft, R. A. C., Dalton, G. B., Efstathiou,
             G., Sutherland, W. J., \& Maddox, S. J.,  1997,
             MNRAS,  291, 305}


\bibitem{deVegaSanchez96} {de Vega, H.J., Sanchez, N., \& Combes, F.,  1996, Phys.Rev. D, 54, 6008}

\bibitem{deVegaSanchez} {de Vega, H.J., Sanchez, N., \& Combes, F.,
              1996, Nature, 383, 56}

\bibitem{deVegaSanchez98}
{de Vega, H.J., Sanchez, N., \& Combes, F., 1998, ApJ, 500, 8}



\bibitem{Einasto02}
 {Einasto, M., Einasto, J., Tago, E., Andernach, H.,
 Dalton, G. B., \& M{\"u}ller, V., 2002, AJ, 123, 51}

\bibitem{Einasto2000b}
{Einasto, M., Einasto, J., Tago, E., et al.,  2007, ApJ, 123, 51.
  arXiv:astro-phy/0012538}


\bibitem{Goldenfeld}{Goldenfeld, N., 1992,
         Lectures on Phase Transitions and Renormalization Group,
         Addison-Wesley Publishing Company}

\bibitem{Gonzalez}
{Gonzalez, A. H., Zaritsky, D., \& Wechler, R. H., 2002, ApJ, 571, 129}


\bibitem{groth1977large}
{Groth, E., \& Peebles, P.J.E., 1977, ApJ, 217, 385}



\bibitem{Hamana}
    Hamana, T. ,  Colombi,  S.,  \&   Suto, Y.,
            2001,      A\&A 367, 18

\bibitem{hawkins20032df}
{Hawkins, E., Maddox, S., Cole, S.,  et al., 2003, MNRAS, 346, 78}


\bibitem{Kaiser1984}
{Kaiser, N., 1984, ApJ, 284, L9}


\bibitem{Kirkwood}
{ Kirkwood, J.G. 1932, J. Chem. Phys., 3, 300}


\bibitem{KlypinKopylov}
{ Klypin, A.A., \& Kopylov, A.I.,
                     1983, Sov. Astron. Lett. ,9, 41.}



\bibitem{loveday1996stromlo}
{Loveday, J.,  Peterson, B.A., Maddox, S.J., \& Efstathiou, G., 1996,  ApJS, 107, 201}


\bibitem{Masters}
{Masters, K. L., Springob, C. M., Haynes, M. P., \& Giovanelli, R., 2006, ApJ, 653, 861}

\bibitem{Padilla03}
{Padilla, N.D., \& Baugh, C.M., 2003, MNRAS, 343, 796}

\bibitem{Peacock2001}
{Peacock, J., Cole, S., Norberg, P., et al., 2001, Nature, 410, 169}

\bibitem{peacockbook}
{Peacock, J. A., 1999,  Cosmological Physics.
Cambridge  Univ. Press, Cambridge, UK}

\bibitem{Peebles}
{Peebles, P. J. E., 1980,
         The Large-scale Structure of the Universe.
            Princeton Univ. Press, Princeton, NJ}


\bibitem{Saslaw85}
{Saslaw, W.C., 1985, Gravitational Physics of Steller
    and Galactic Systems. Cambridge University Press.}

\bibitem{Saslaw2000}
{Saslaw, W.C.,  2000,  The Distribution of the Galaxies.  Cambridge University Press.}


\bibitem{Schwinger}
{Schwinger, J., 1951, Proc. Natl. Acamd. Sci. USA ,37, 452, 455.}

\bibitem{Shaver}
{Shaver, P., 1988,
   in Large-Scale Structure of the Universe,
        IAU Symp. 130, ed. J. Audouze et al. ,Dordrecht:  Reidel, 359}

\bibitem{SoneraPeebles}
{Soneira, R.M., \& Peebles, P.J.E. 1987, AJ , 83, 845}


\bibitem{SzalaySchramm}
{Szalay, A., \& Schramm, D., 1985, Nature,  314, 718.}

\bibitem{Taruya2001}   Taruya, A., Magira, H., Jing, Y.P., \& Suto, Y.,
    2001,     Publ Astron Soc Jpn    53  (2):  155.


\bibitem{TotsujiKihara}
  {Totsuji, H., \& Kihara, T., 1969, PASJ, 21, 221}


\bibitem{Yahata}
Yahata, K., Suto, Y., Kayo, I., et al.,  2005, Publ Astron Soc Jpn   57  (4):  529.

\bibitem{Yoshikawa} Yoshikawa, K., Taruya, A., Jing, Y.P., \& Suto, Y.,
                  2001,    Astrophys.J. 558,  520

\bibitem{Zandivarez}
{Zandivarez, A., Merchan, M. E., \& Padilla, N. D., 2003, MNRAS, 344, 247}

\bibitem{Zehavi05}
{Zehavi, I., Zheng, Z., \& Weinberg, D. H., et al. 2005, ApJ, 630, 1}

\bibitem{Zhang2007}
{Zhang, Y., 2007, A\&A,  464, 811}

\bibitem{zhang2009nonlinear}
{Zhang, Y., \& Miao, H. X., 2009, Research in Astron. and Astrophys., 9, 501}



\end{thebibliography}
\end{document}